\documentclass[showpacs,preprintnumbers,aps,prb,preprint]{revtex4}

\usepackage[dvips,xdvi]{graphicx}
\usepackage{epsfig}

\begin {document}
\title{Temperature dependence of exciton
Auger decay process in Cuprous Oxide}
\author{Yingmei Liu, David Snoke}
\affiliation{Department of Physics and Astronomy\\University of
Pittsburgh, Pittsburgh, PA 15260}
\date{}
\begin{abstract}
We report theoretical and experimental study on Auger decay of
excitons in Cu$_{2}$O in a wide range of temperatures, from 2
Kelvin to 325 Kelvin, in single-photon surface excitation. We find
that the Auger constant linearly increases with temperature above
100 Kelvin. This constant is important for experiments on Bose
condensation of excitons.
\end{abstract}
\maketitle

\newpage
\section{INTRODUCTION}
There is a long history in investigating excitonic effects in
naturally-grown high-quality Cu$_2$O, because of the inversion
symmetry in Cu$_2$O and the large binding energy of the excitons,
150 meV. The inversion symmetry forbids any direct dipole
recombination of excitons which gives excitons long lifetime up to
microsecond, and 150 meV is equivalent to a temperature of 1740
Kelvin which allows excitons to exist at room temperature
\cite{auger}. We are only interested in the lowest excitonic
state, the ``yellow" exciton series, consisting of electrons in
the $\Gamma_6^+$ level of the conduction band and holes in the
$\Gamma_7^+$ level of the valence band with 2.173 eV gap energy.
By electron-hole exchange, the ``yellow" excitons split into a
triplet orthoexciton and a singlet paraexciton, which lies 12 meV
lower \cite{auger}.

In the last ten years, there have been a lot of experiments done
with excitons in Cu$_2$O, either with an intense surface
excitation which creates a highly non-equilibrium exciton gas, or
with an inhomogeneous strain which confines excitons in a harmonic
potential trap with almost constant volume. One interesting topic
among those experiments is the importance of the Auger
non-radiative recombination process in Bose condensation of
excitons. In the Auger process, two excitons collide, and one
exciton ionizes, taking the energy of the other exciton which
recombined. Some groups believe that the density-dependent Auger
process is a barrier for Bose condensation of excitons, since it
not only gives a severe limit in exciton density, but also heats
up the exciton gas \cite{auger,warren,O'Hara,kavoulakis}, while
others claim that the process is negligible \cite{jolk}.
Trauernicht \textit{et al.} first gave strong evidences of an
Auger process by trapping the excitons with an external stress
\cite{trauernicht}.

Our previous work \cite{auger} has shown that the Auger
recombination process is important and increases roughly as the
square of the applied stress, but it left one question in doubt,
why there is a very small Auger effect at stresses lower than 1.5
kbar at 2 Kelvin, while a strong Auger effect must occur in
surface excitation with a negligible stress at higher
temperatures, for example at room temperature. Does the surface
play an important role in the Auger effect, or is the Auger effect
temperature dependent? In this work, our theoretical and
experimental results show that the Auger decay rate increases
linearly with temperature, which answers the above question.
\section{EXPERIMENT}
Excitons are created at one [110] surface of a $4.9\times3\times3$
mm naturally-grown, high-quality Cu$_2$O crystal with
single-photon near-resonant excitation under negligible applied
stress in the high temperature region, 115$\sim$325 Kelvin. An
imaging lens collects exciton luminescence from the same [110]
surface and focuses the light onto an entrance slit of a
$\frac{1}{4}$-meter spectrometer which is connected with a CCD
camera and a photomultiplier (PMT). The exciton population
dynamics are examined by means of time-correlated single photon
detection. This experimental approach has several advantages.
First, in the high temperature region, paraexciton-to-orthoexciton
conversion leads the excitons to reach thermal equilibrium
quickly. Second, as pointed out by Warren and Wolfe \cite{warren},
the diffusion constant at high temperature is much smaller than
that at low temperature, which allows a constant-volume
measurement. Also, in these experiments, we use a cavity-dumped
ultra-fast dye laser which has a long period, 260 ns between
pulses. Therefore paraexcitons definitely reach thermal
equilibrium with orthoexcitons. The bath temperature in the
experiments is controlled by a homemade PID controller with an
accuracy of $\pm$ 2 Kelvin.

In Cu$_2$O, there are two possible single-photon transitions, a
direct transition by emitting a photon, and a phonon-assisted
transition by emitting a photon and an optical phonon. Because the
indirect transition is allowed for all exciton states, the
indirect phonon-assisted luminescence spectrum shows the energy
distribution of excitons \cite{negoita}. At high temperature, the
direct single-photon transitions for orthoexcitons and
paraexcitons are almost forbidden, while the phonon-assisted
process depends weakly on temperature \cite{snoke}. In Section 4,
we will talk about phonon-assisted orthoexciton recombination
processes, assisted by all possible optical phonons and acoustic
phonons, although the orthoexciton creation process via a
$\Gamma_{12}^-$ longitudinal optical phonon is much stronger than
other phonon-assisted orthoexciton creation processes
\cite{snoke}.

In the surface excitation, the absorption coefficient of Cu$_2$O
has a strong relation to bath temperature and laser photon energy
\cite{absorption, grun}. We calculate absorption lengths of
Cu$_2$O for the input laser at 606 nm and for the orthoexciton
luminescence at each temperature from Pastrnyak's report
\cite{absorption} and J. B. Grun's paper \cite{grun} on the
absorption coefficient of Cu$_2$O, as shown in Figure 1. With a
focused laser spot with radius of 120 $\mu$m, the orthoexciton has
a volume of $\pi\times$(120
$\mu$m)$^{2}\times$(Absorption-length), which is in the order of
$10^{-6}cm^{-3}$. As we did in previous works \cite{auger, snoke},
we assume that one absorbed photon produces an exciton, since the
laser is tuned close to the exciton resonance. At the highest
laser power of 2.64 nJ per pulse, with 30$\%$ absorbed in the
crystal after considering all the optics involved, there are
10$^{9}$ orthoexcitons created initially, which implies an initial
orthoexciton density in the order of $10^{15} \mbox{cm}^{-3}$.

Because the absorption length at the orthoexciton luminescence
line is of the order of a millimeter at high temperatures, which
is comparable to the width of Cu$_{2}$O sample, we have to add a
correction factor to deduce the exciton density from the
luminescence intensity at each temperature. The correction factor,
CF, is equal to the ratio between $I_{detected}$, the real
detected luminescence intensity, and $I_{0}$, the ideal detected
luminescence intensity without taking into account of the
absorption of the luminescence in Cu$_{2}$O, by a Photomultiplier.

If we assume that one absorbed photon creates one exciton, the
total number of excitons at a certain position $x$ in the sample,
is proportional to the number of absorbed photons,
$n_x\propto(e^{\frac{-x}{L_1}})$, where $0\leq x\leq L$ and L, the
width of Cu$_{2}$O, is 3 millimeter. L$_{1}$ and L$_{2}$ are the
absorption lengths for the laser and orthoexciton luminescence at
each temperature, respectively. Because of the absorption of the
exciton luminescence in the sample, the number of the detected
exciton from the position $x$ is proportional to
$(e^{\frac{-x}{L_2}})$. In other words, the total contribution to
the luminescence from the position $x$ is proportional to
$(e^{\frac{-x}{L_1}}e^{\frac{-x}{L_2}})$.

Therefore, the detected luminescence intensity is proportional to
the integration of the detected excitons over the whole length of
the sample, and the correction factor is calculated as the
following,
\begin{eqnarray}
CF=\frac{I_{detected}}{I_{0}}=\int^{L}_{0}e^{\frac{-x}{L_{1}}}e^{\frac{-x}{L_{2}}} d x\quad\div\quad\int^{L}_{0}e^{\frac{-x}{L_{1}}} d x\\
\Rightarrow CF=\frac{1}{L_{1}}\cdot
\frac{1}{\frac{1}{L_{1}}+\frac{1}{L_{2}}}\cdot\frac{1-e^{-L(\frac{1}{L_{1}}+\frac{1}{L_{2}})}}{1-e^{-\frac{L}{L_{1}}}}
\end{eqnarray}

In principle, we should account for re-emission from the
reabsorbed photons, but since the re-emission mostly goes in
directions other than x (see inset of Figure 1), this will be a
small correction.

\section{ RESULTS AND DISCUSSIONS}
Figure 2 shows the typical orthoexciton photoluminescence
intensity as a function of time after a short laser pulse, at two
different temperatures, 150 and 325 Kelvin, and under three
different pulse energies, 2.64 nJ, 2.64/2=1.32 nJ and 2.64/6=0.44
nJ, multiplied by overall factors of 1, 2 and 6 respectively. At
each temperature, the different decay times for different laser
powers indicate that there is the density-dependent Auger process
of excitons.

Our previous work \cite{auger, snoke} presented coupled rate
equations including the Auger decay rate, which explained well the
data at 2 Kelvin under different stresses and laser pulse
energies. However, because of phonon emission and absorption, the
orthoexciton and paraexciton are well coupled in the high
temperature region \cite{auger}, and therefore the coupled rate
equations can be modified to the following:
\begin{eqnarray}
\frac{d n_{o}}{d t}  & = &
-\frac{n_{o}}{\tau_{o-p}}-A_{o}n_{o}^{2}+\frac{3}{8}(A_{o}n_{o}^{2}+A_{p}n_{p}^{2})-\frac{n_{o}}{\tau_{o}}+\frac{n_{p}}{\tau_{p-o}}\quad,\\
\frac{d n_{p}}{d t}  & = &
\frac{n_{o}}{\tau_{o-p}}-A_{p}n_{p}^{2}+\frac{1}{8}(A_{o}n_{o}^{2}+A_{p}n_{p}^{2})-\frac{n_{p}}{\tau_{p}}-\frac{n_{p}}{\tau_{p-o}}\quad.
\end{eqnarray}
In these equations, indies o and p stand for ``orthoexciton" and
``paraexciton", respectively, $A$ is the Auger decay constant, n
is the exciton density, $\tau$ is the decay time or conversion
time, and the indies o-p and p-o mean orthoexciton-paraexciton
conversion and paraexciton-orthoexciton conversion, respectively.
In each Auger process, two excitons collide and end up with one
exciton recombining and the other exciton ionizing. Therefore,
only the ionized exciton, half of the excitons participating in
the Auger process, will be added back to exciton population.
Because the orthoexciton is a triplet state and the paraexciton is
a singlet state and spin is randomly selected in ionization,
$\frac{3}{4}$ of the ionized excitons in the Auger process will be
returned as orthoexcitons and $\frac{1}{4}$ of these excitons will
be returned as paraexcitons. We have to consider inter-conversions
between paraexcitons and orthoexcitons in the high temperature
region, although only orthoexcitons are converted to paraexcitons
at 2 Kelvin \cite{kavoulakis}. Because energy splitting between
orthoexciton and paraexciton at the zone center is 12 meV, and
orthoexciton and paraexciton reach thermal equilibrium rapidly at
high temperature \cite{warren}, the conversion time in the
inter-conversion mechanisms has a relationship,
\begin{eqnarray}
\tau_{p-o}=\tau_{o-p}\times e^{\frac{12 meV}{K_{B}T}}\quad.
\end{eqnarray}

All the data, under ten different temperatures and three different
laser pulse energies at each temperature, are well fit with the
above coupled equations, as shown on the solid lines of Figure 2
(a) and (b). The fit results imply that the Auger constant of the
orthoexciton is equal to that of the paraexciton at each
temperature. The results also show that the radiative lifetimes of
paraexcitons and orthoexcitons are on the order of several hundred
nanoseconds, and the inter-conversion time between paraexcitons
and orthoexcitons is on the order of 0.1 nanosecond which is
consistent with the report from Wolfe's group \cite{warren} and
our previous results \cite{auger}. Because paraexcitons rapidly
reach thermal equilibrium with orthoexcitons, in the fitting
process we can assume that at all times after t = 0.1 ns,
\begin{equation}
n_{p}(t) \quad= \quad n_{o}(t)\times
e^{\frac{12meV}{K_{B}T}}\quad.
\end{equation}

An interesting result from the fitting is that the Auger constant
of excitons linearly increases with temperature in the region from
115 Kelvin to 325 Kelvin, as shown in Figure 3. Several groups
have studied the temperature dependence of the Auger decay process
in Cu$_2$O, but the highest temperature they used is 77 Kelvin
\cite{warren, lowT}. Their works show that the Auger constant of
orthoexcitons is almost independent of temperature. The above two
results are not inconsistent, which is explained in detail in next
section.

\section{ Phonon-assisted mechanism }
Kavoulakis and Baym \cite{lowT} introduced a model based on
Fermi's Golden rule and \textbf{k$\cdot$p} theory to understand
the Auger recombination process with an optical phonon-assisted
mechanism at low temperature. We follow their steps, but modify it
with considerations of high temperature effects and all possible
phonons in an Auger process, in which two excitons with momentum
\textbf{$\vec{K}$} and \textbf{$\vec{P}$} collide, producing one
ionized electron-hole pair with momenta \textbf{$\vec{k}$$_e$} and
\textbf{$\vec{k}$$_h$}, and absorbing or emitting a phonon.

In the phonon-assisted process, $\frac{\Gamma_{Auger}}{V}$, the
Auger decay rate per unit volume is a function of $\Gamma_{K,P}$
and $f_{K}$, $f_{P}$, the decay rate and distribution functions of
the two excitons, respectively:
\begin{equation}
\frac{\Gamma_{Auger}}{V}=
\frac{1}{V}\sum_{K,P}f_{K}f_{P}\Gamma_{K,P}\sim A n^{2}\quad,
\end{equation}
Where A is Auger constant for the exction and n is the density of
the excitons.

From Fermi's Golden Rule, the decay rate of two excitons with
momenta \textbf{$\vec{K}$} and \textbf{$\vec{P}$} is
\begin{eqnarray}
\Gamma_{K,P}&=&\frac{2\pi}{\hbar}\sum_{k_{e},k_{h},Q}|M|^{2}(1-n_{c,k_{e}})(1-n_{v,k_{h}})\nonumber\\
&&\times[(n_{ph}+1)\delta(E_{K}+E_{P}-\hbar\omega_{Q}-\varepsilon_{c,k_{e}}-\varepsilon_{v,k_{h}})\nonumber\\
&&+n_{ph}\delta(E_{K}+E_{P}+\hbar\omega_{Q}-\varepsilon_{c,k_{e}}-\varepsilon_{v,k_{h}})]\quad,
\end{eqnarray}
The first term in Equation 8 is for a process emitting a phonon
with momentum \textbf{$\vec{Q}$}, while the second term is for a
phonon absorption process by absorbing a phonon with momentum
\textbf{$\vec{Q}$}. The term $n_{ph}=1/(e^{E_{ph}/(k_{B}T)}-1)$ is
the density of the phonon at temperature T, and $k_{B}$ is
Boltzmann's constant.

In the equations, based on \textbf{k$\cdot$p} theory \cite{lowT},
the matrix element of an optical phonon-assisted process,
M$_{optical}$, is given by the following equation,
\begin{eqnarray}
M_{optical}\quad\approx\quad2^{7}\pi\frac{e^{2}}{\epsilon_{\infty}a_{B}}\frac{\hbar}{m}\times\frac{|\textbf{P}_{v,c'}|}{(\varepsilon_{c',0}-\varepsilon_{v,0})}\times\frac{k_{e}a_{B}}{[1+(k_{e}a_{B})^{2}]^{3}}\nonumber\\
\quad\times(\frac{\hbar^{2}D_{optical}^{2}}{2V\rho
E_{optical}})^{\frac{1}{2}}\frac{1}{(\varepsilon_{c,0}-\varepsilon_{c',0})}\times\delta_{\textbf{$\vec{k}$}_{e}+\textbf{$\vec{k}$}_{h}\pm
\textbf{$\vec{Q}$},\textbf{$\vec{K}$}+\textbf{$\vec{P}$}}\quad.
\end{eqnarray}
where D$_{optical}$ is the deformation potential for an optical
phonon, $E_{optical}$ is the energy of the optical phonon,
$\varepsilon_{c}$ and $\varepsilon_{c'}$ are energies of the
lowest conduction band $\Gamma_6^+$ and the only odd parity
conduction band $\Gamma_8^-$, respectively, $\varepsilon_{v}$ is
the energy of the valence band $\Gamma_7^+$, $P_{v,c'}$ is the
matrix element of the momentum operator between the conduction and
valence band at zone center, $\rho$ is the mass density of
Cu$_2$O, $\epsilon_{\infty}$ is the dielectric constant, and
a$_{B}$ is the exciton Bohr radius.

In an acoustic phonon assisted process, $k_{necessary}$ the
necessary momentum for an acoustic phonon is to satisfy with
energy and momentum conservations between the lowest conduction
band and the only odd parity conduction band, that is to say,
$k_{necessary}$ = $(\varepsilon_{c}-\varepsilon_{c'})/(\hbar\nu)$
= 1.515$\times 10^{9}$ cm$^{-1}$. In the discussion, $\nu$ is the
acoustic phonon speed. However, $k_{max}$, the maximum momentum of
an acoustic phonon, is equal to $\frac{\pi}{a_{L}}$ = 7.375$\times
10^{7}$ cm$^{-1}$ because it is restricted to be in the first
Brillouin zone, where $a_{L}$ is the lattice constant of
Cu$_{2}$O. Therefore there is no possible process assisted by the
acoustic phonon and from now on all discussions will only consider
the process assisted by the optical phonon.

As mentioned by Kavoulakis \cite{lowT}, $E_{K}$, $E_{k_{e}}$, and
$E_{k_{h}}$, the energy of an exciton with momentum
$\textbf{$\vec{K}$}$, the energy of an electron in the conduction
band, and the energy of a hole in the valence band, respectively,
are given by \cite{lowT},
\begin{eqnarray}
E_{K}&=&E_{gap}+\frac{\hbar^{2}K^{2}}{2m_{exciton}}\\
E_{k_{e}}&=&E_{gap}+\frac{\hbar^{2}k_{e}^{2}}{2m_{e}}\\
E_{k_{h}}&=&\frac{\hbar^{2}k_{h}^{2}}{2m_{h}}
\end{eqnarray}
Because the momenta of excitons, $\textbf{$\vec{K}$}$ and
$\textbf{$\vec{P}$}$, are determined by exciton thermal motion, we
can make a reasonable assumption \cite{lowT},
\begin{equation}
 \textbf{$\vec{K}$},\textbf{$\vec{P}$}\ll\textbf{$\vec{k}$$_{e}$},\textbf{$\vec{k}$$_{h}$}\quad.
\end{equation}

By substituting Equation (9) through (13) into Equation (8),
$\Gamma_{K,P}$ the decay rate of two excitons in the
phonon-assisted process is equal to
\begin{eqnarray}
\Gamma_{K,P}^{phonon-assisted}&=&\frac{2^{14}\pi^{3}e^{4}\mu\hbar}{\epsilon_{\infty}^{2}a_{B}^{2}Vm^{2}E_{optical}\rho}\times\frac{D_{optical}^{2}}{(\varepsilon_{c,0}-\varepsilon_{c',0})^{2}}\times\frac{|\textbf{P}_{c',v}|^{2}}{(\varepsilon_{c',0}-\varepsilon_{v,0})^{2}}\nonumber\\
&&\times
\frac{(k_{e}a_{B})^{2}}{[1+(k_{e}a_{B})^{2}]^{6}}\times[1+\frac{2}{e^{E_{optical}/k_{B}T}-1}]\quad,
\end{eqnarray}

The integration is restricted to the first Brillouin zone. All
constant parameters used in the calculation are listed in Table 1.
The optical deformation potentials and $|\textbf{P}_{c,v}|^{2}$
are estimated from fitting of experimental data.

Therefore, by substituting Equation (14) into Equation (7), we get
$A_{phonon-assisted}$, the phonon-assisted Auger constant, in the
unit of cm$^3$/ns,
\begin{eqnarray}
A_{phonon-assisted}\simeq
3.698\times10^{-17}\times[1+\frac{2}{e^{160/T}-1}]+1.298\times10^{-17}\times[1+\frac{2}{e^{217/T}-1}].
\end{eqnarray}
Only two optical phonons, $\Gamma_{12}^{-}$ and $\Gamma^{-}_{15}$,
make important contributions.

We plot Equation 15, a theoretical prediction of the exciton Auger
constant assisted by all possible phonons from 2 to 325 Kelvin,
and our experimental Auger recombination constants in the high
temperature region ( 115 $\sim$ 325 Kelvin ) in Figure 3, which
shows that our experimental data in the high temperature region
can be well fit by the theoretical prediction of the Auger
constant from the phonon-assisted mechanism.

At low temperatures, where the phonon energy is much smaller than
the thermal energy k$_{B}$T, n$_{ph}$ the density of the phonon is
much less than 1. Therefore only the phonon emission process is
important and phonon absorption process is negligible. We can
approximate:
\begin{eqnarray}
\Gamma_{K,P}&=&\frac{2\pi}{\hbar}\sum_{k_{e},k_{h},Q}|M_{ph}^{2}|(1-n_{c,k_{e}})(1-n_{v,k_{h}})\nonumber\\
&&\times
\quad\delta(E_{K}+E_{P}-\hbar\omega_{Q}-\varepsilon_{c,k_{e}}-\varepsilon_{v,k_{h}})\quad\quad(
\mbox{Low T} )
\end{eqnarray}
After substituting Equation (16) into Equation (7), we find that
the Auger decay constant is weakly dependent on temperature when
the temperature is lower than 77 Kelvin, which is shown in Fig. 3
and also consistent with the reports from Warren \emph{et al.}
\cite{warren} and Kavoulakis \emph{et al.} \cite{warren,lowT}.

Therefore, we can explicitly state our conclusion on the
temperature dependence of the Auger recombination constant of
excitons as the following: the Auger constant is weakly dependent
on temperature when the temperature is below 77 Kelvin, and
linearly increases with temperature in the high temperature
region.

To verify our model, we make two comparisons. The first comparison
is between a reported Auger constant from Wolfe's group
\cite{warren} and our theoretical prediction at 77 Kelvin, which
are consistent with each other, though our theoretical prediction
is about 4 percent larger, as shown in Figure 4. The other
comparison is made between the experimental data and our
theoretical model at 2 Kelvin. Our theoretical result indicates
that the Auger constant does not continuously decrease to zero at
zero stress at 2 Kelvin, but becomes a constant,
$4.996\times10^{-17}$ cm$^{3}/$ns, which is comparable to the
experimental data at 1.5 kbar at the same temperature reported in
the previous work \cite{auger}, as shown in Fig. 4 also. The
comparison implies several things. First, the Auger constant has a
negligible stress dependence when the stress is lower than 1.5
kbar. Second, the two results are consistent, although the
theoretical prediction from our phonon-assisted mechanism is about
9 percent larger than the extrapolated data. Third, the surface
doesn't play an important role in the Auger effect. Since the 4 to
9 percent discrepancy between experimental data and theoretical
prediction is in the acceptable experimental error range, our
modified phonon-assisted mechanism may be a good model for the
Auger recombination rate of excitons in both low and high
temperature regions.
\section{CONCLUSIONS}
From the above theoretical and experimental studies, we may say
that in surface excitations, the Auger recombination constant
linearly increases with temperature at high temperatures, and
weakly depends on temperature in the low temperature region, which
is consistent with our modified model of phonon-assisted
Auger recombination. However, we do not have a model for the stress dependence.\\

{\bf Acknowledgements}. This work has been supported by CRDF-MRDA
Award No. MP2-3026 and NSF Award No. DMR-0102457. We thank S.
Denev for his contribution on the PID temperature controller.
Samples of Cu$_2$O was obtained from P. J. Dunn of the Smithsonian
Institute.

\newpage

\begin{table}[!h]
\begin{center}
\begin{tabular}{|c|c|c|}
\hline Parameter & Value& References\\
\hline $\epsilon_{\infty}$&6.46&11\\
\hline $\rho$&6 gram/cm$^{3}$&13\\
\hline $\nu$&4.5$\times 10^{5}$ cm/second&12\\
\hline $m_{e}$&$m_{e}^{0}$&4\\
\hline $m_{h}$&0.7$m_{e}^{0}$&4\\
\hline $m_{e}^{0}$&9.1$\times10^{28}$ gram&14\\
\hline
$a_{B}$&7$\times10^{-8}$ cm&10\\
\hline
$a_{L}$&4.26$\times10^{-8}$ cm&10\\
\hline $D_{\Gamma_{12}^{-}}$&1.10$\pm$0.01 eV/\AA&this
work\\
\hline
$D_{\Gamma_{15}^{-}}$&0.75$\pm$0.01 eV/\AA&this work\\
\hline
$|\textbf{P}_{c',v}|/\hbar$&0.55$\pm$0.01 \AA$^{-1}$&this work\\
\hline
$\varepsilon_{c'}-\varepsilon_{c}$&0.449 eV&11\\
\hline
$\varepsilon_{c'}-\varepsilon_{v}$&2.622 eV&11\\
\hline
$\Gamma_{12}^{-}$&13.8 meV&15\\
\hline
$\Gamma_{15}^{-}$&18.7 meV&15\\
\hline
\end{tabular}
\caption{\small{Constant parameters used in the Auger decay rate
calculation.}}
\end{center}
\end{table}

\begin{figure}[!ht]
\begin{center}
\includegraphics[width=7in, height=5in]{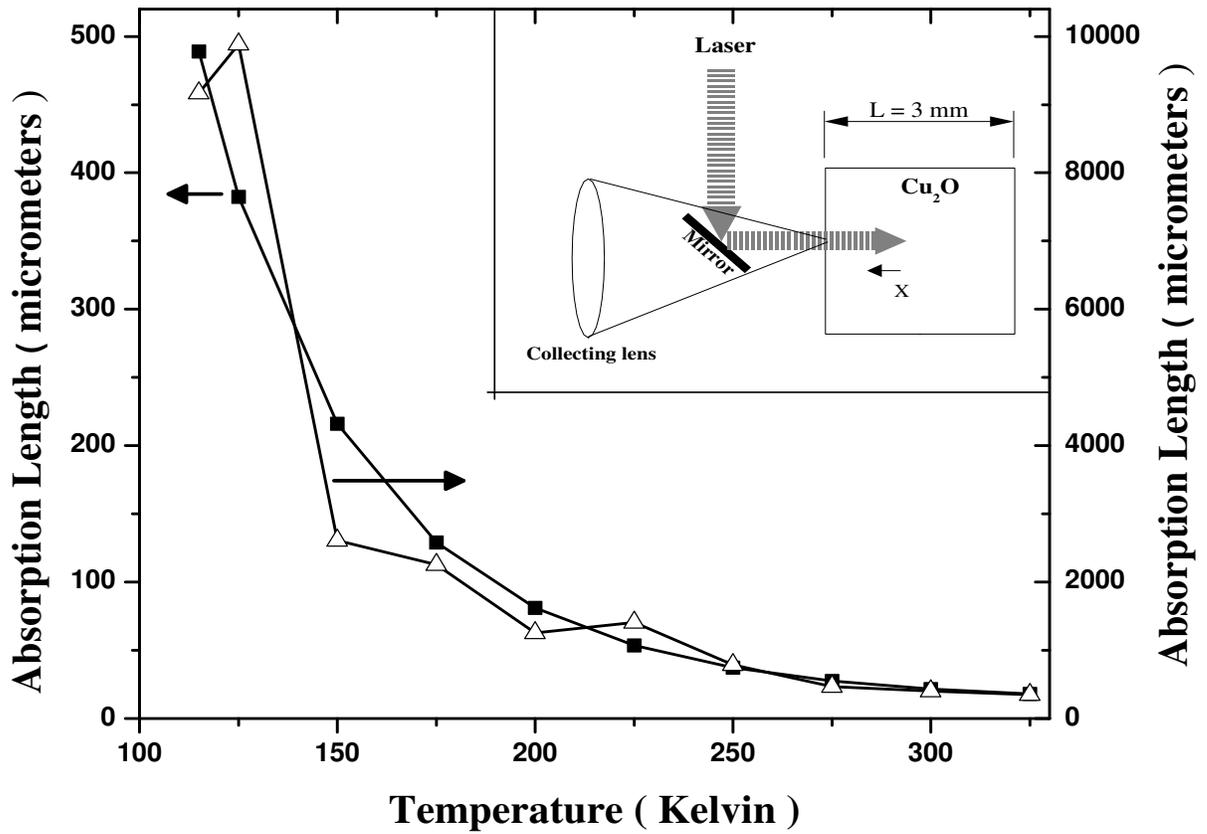}
\caption{\label{fig1.fig} \small{Absorption lengths of the input
laser at 606 nm ( black squares ), and the exciton luminescence (
open triangles ) in Cu$_2$O at each temperature in the high
temperature region. Inset: Schematic of the experimental set-up in
this work.}}
\end{center}
\end{figure}

\begin{figure}[!ht]
\begin{center}
\includegraphics[width=5in,height=7in]{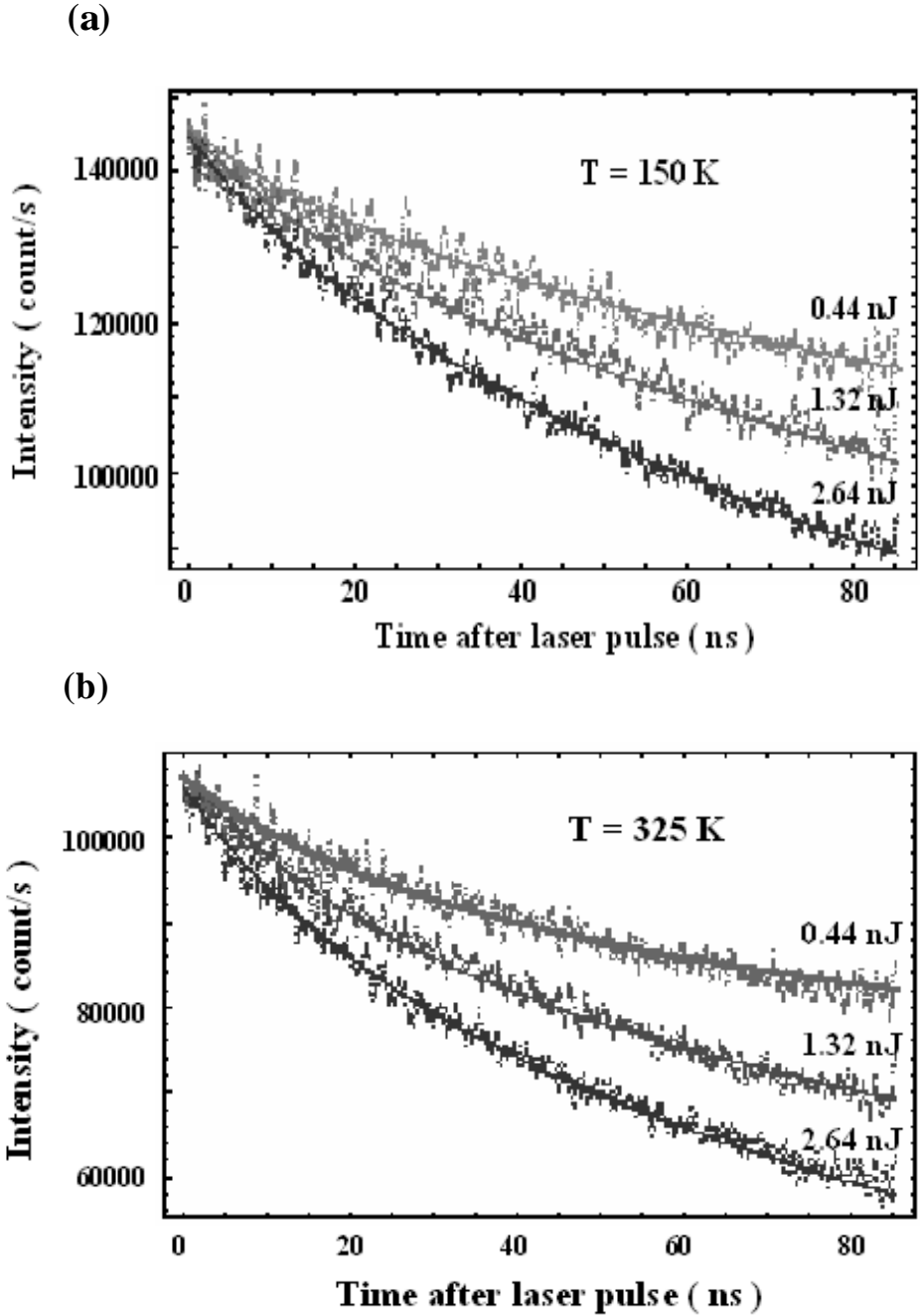}
\caption{\label{fig2(ab).fig} \small{(a) Orthoexciton luminescence
decay at T = 150 Kelvin for three different pulse energies, 2.64
nJ, 1.32 nJ and 0.44 nJ, as shown in the figure. Dots:
Phonon-assisted orthoexciton luminescence intensity as a function
of time after a short laser pulse. Solid lines: fit to the coupled
rate equations discussed in the text. (b) Orthoexciton
luminescence decay, as in (a), but at T = 325 Kelvin.}}
\end{center}
\end{figure}

\begin{figure}[!ht]
\begin{center}
\includegraphics[width=7in,height=5in]{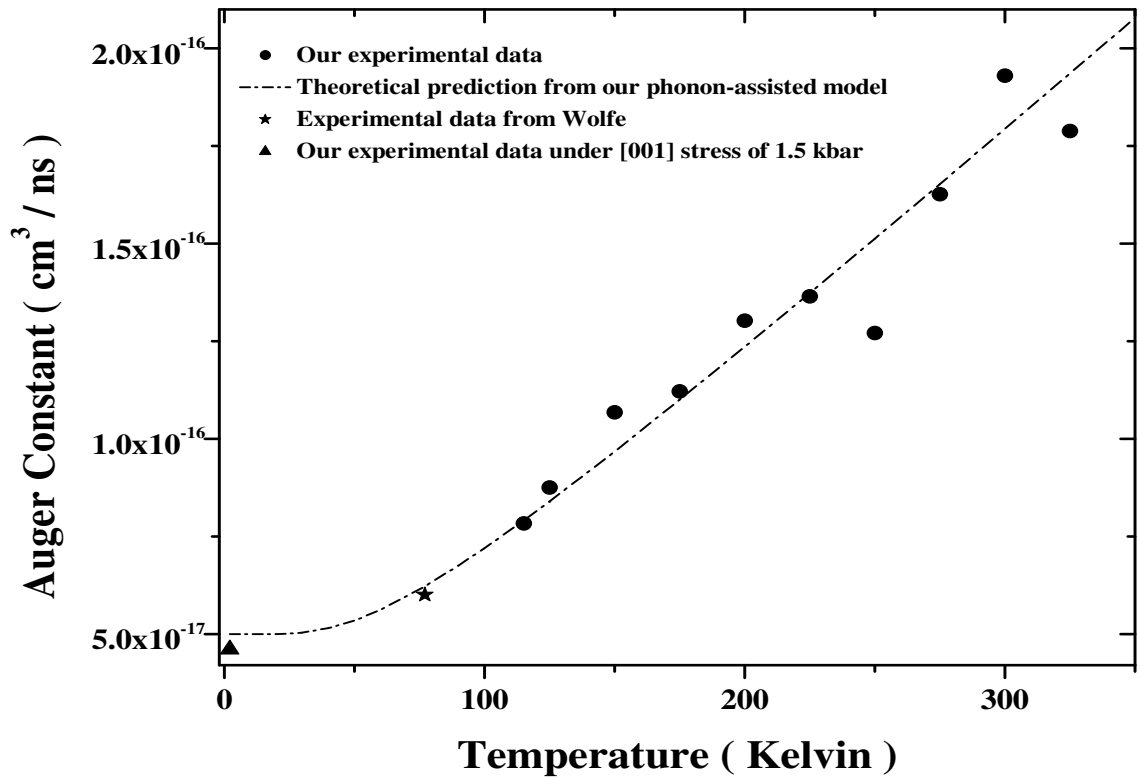}
\caption{\label{fig3.fig} \small{Predicted and experimental Auger
constants of the excitons as a function of temperature in the high
and low temperature regions.}}
\end{center}
\end{figure}

\begin{figure}[!ht]
\begin{center}
\includegraphics[width=7in,height=5in]{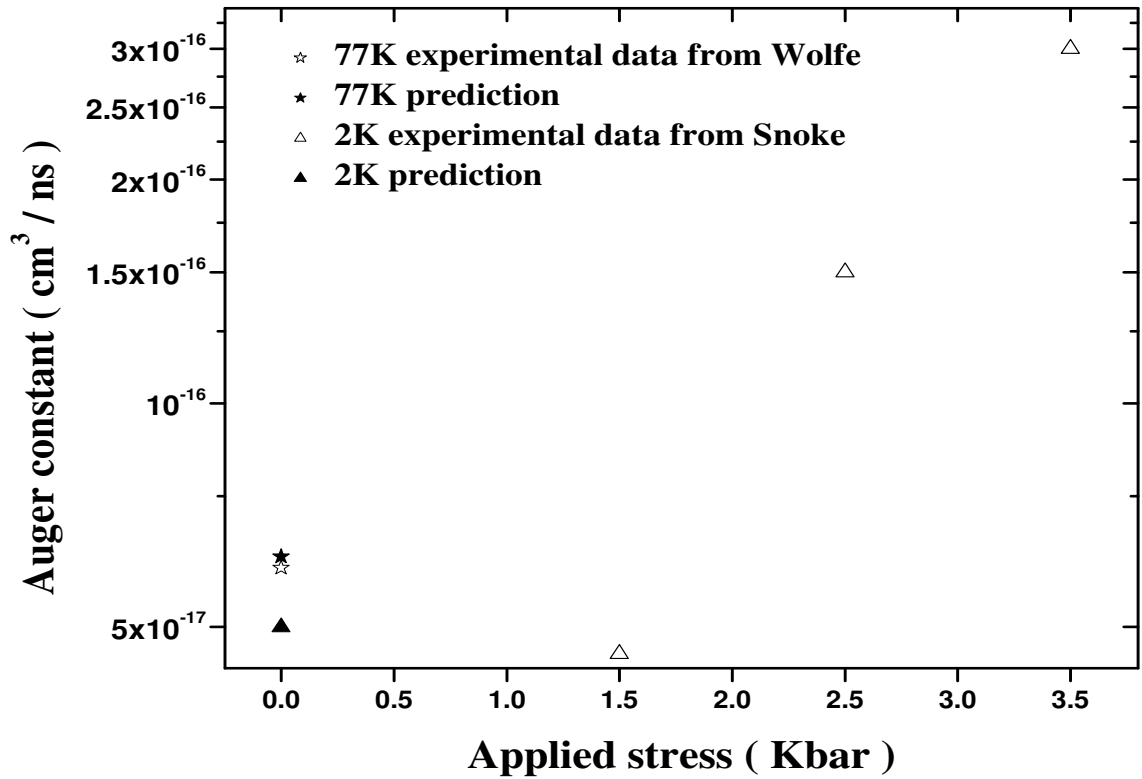}
\caption{\label{fig4.fig} \small{Stress and temperature dependence
of the orthoexciton Auger constant. Open triangles: data reported
by Snoke \emph{et al.} \cite{auger} at 2 Kelvin for three
different stresses, 1.5, 2.5 and 3.5 kbar. Open stars: data
reported by Wolfe \emph{et al.} \cite{warren} with negligible
stress at 77 Kelvin. Black triangles and black stars are
predictions from our theoretical model with negligible stress at 2
Kelvin and 77 Kelvin, respectively.}}
\end{center}
\end{figure}

\end {document}